\documentclass[twocolumn,prl,showpacs,preprintnumbers,amsmath,amssymb]{revtex4}

\usepackage{graphicx}
\usepackage{subfigure}

\begin{document}

\title{Organometallic Benzene-Vanadium Wire: \\One-Dimensional Half-Metallic Ferromagnet.}

\author{
Volodymyr V.~Maslyuk$^{1}$, Alexei Bagrets$^{1,2}$, Thomas
Bredow$^{3}$ and Ingrid Mertig$^{1}$ }

\email{volodymyr@physik.uni-halle.de}%

\affiliation{
$^1$Martin-Luther-Universit\"at Halle-Wittenberg, Fachbereich Physik, D-06099 Halle, Germany \\
$^2$Institut f{\"u}r Nanotechnologie, Forschungszentrum Karlsruhe, D-76344, Germany \\
$^3$Institut f{\"u}r Theoretische Chemie, Universit\"at Hannover, D-30167 Hannover, Germany}




\date{\today}

\begin{abstract}
Using density functional theory we have performed theoretical
investigations of the electronic properties of a free-standing
one-dimensional organometallic vanadium-benzene wire. This
system represents the limiting case of multi-decker
V$_n$(C$_6$H$_6$)$_{n+1}$ clusters which can be synthesized.
We predict that the ground state of the wire is a 100\%
spin-polarized ferromagnet (half-metal). Its density of states is
metallic at the Fermi energy for the minority electrons and shows
a semiconductor gap for the majority electrons. We found that the
half-metallic behavior is conserved up to 12\%, longitudinal
elongation of the wire. However, under further stretching, the
system exhibits a transition to a high-spin ferromagnetic state
that is accompanied by an abrupt jump of the magnetic moment and a
gain of exchange energy.
\end{abstract}


\maketitle

During the last years, molecular magnets have been attracting
enormous attention, because they are considered as
potential candidates for future applications
in high-density information storage and quantum computing.
Among such novel systems, we would like to focus on
vanadium-benzene multi-decker clusters (V$_n$Bz$_{n+1}$,
Bz=C$_6$H$_6$). These are molecular magnets appearing in
a gas phase \cite{1,2}. The synthesis of the V-Bz clusters
was realized during a reaction of
laser-vaporized metal atoms with benzene in He atmosphere
\cite{2,3}. The mass spectrometric measurements \cite{2,4} and
theoretical studies \cite{5,6,7} suggested that V$_n$Bz$_{n+1}$
clusters, approaching a length of up to $n=6$ (18~\AA), should
have a one-dimensional structure which is settled by the
metal-ligand interactions confined to a single molecular axis.

In this Letter, we present the study of the infinite
one-dimensional (1D) organometallic vanadium-benzene (V-Bz)
wire shown in Fig.1 which could be considered
as the limiting case of the V$_n$Bz$_{n+1}$
clusters with $n \to \infty$. We have performed {\it ab initio}
calculations of the electronic structure of the system and predict
that the wire under consideration is a {\it new}\ unusual example
of the {\it one-dimensional}\ half-metallic ferromagnet. It has
integer magnetic moment of $1\mu_B$ per unit cell and the
spin-polarized band structure with finite density of states at the
Fermi level for one spin channel and a semiconductor gap for the
other spin channel (Fig.2). Thus, the electronic properties of
this organometallic wire are similar to that of the known Heusler
compound NiMnSb which has been predicted by de Groot {\it et al.}\
in 1983 to be a half-metal \cite{deGroot}. Due to a spectacular
electronic structure such wire or its finite fragment would act as
a nearly perfect spin-filter with very high magnetoresistance
ratio when it is placed between ferromagnetic electrodes in the
two-terminal spin-polarized-STM like set-up \cite{STM}.

\begin{figure}[b]
  \centering \includegraphics[height=2.0in]{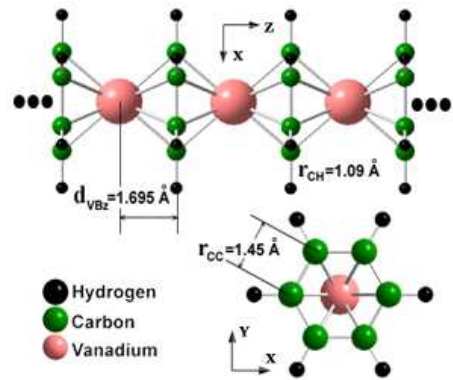}
  \caption{\label{Struct} (Color online) The structure of
the V(C$_6$H$_6$) wire.}
\end{figure}

\begin{figure*}[t]
   \begin{minipage}[lc]{0.78\textwidth}
      \centering \includegraphics[bb=24 20 450 256,height=2.7in]{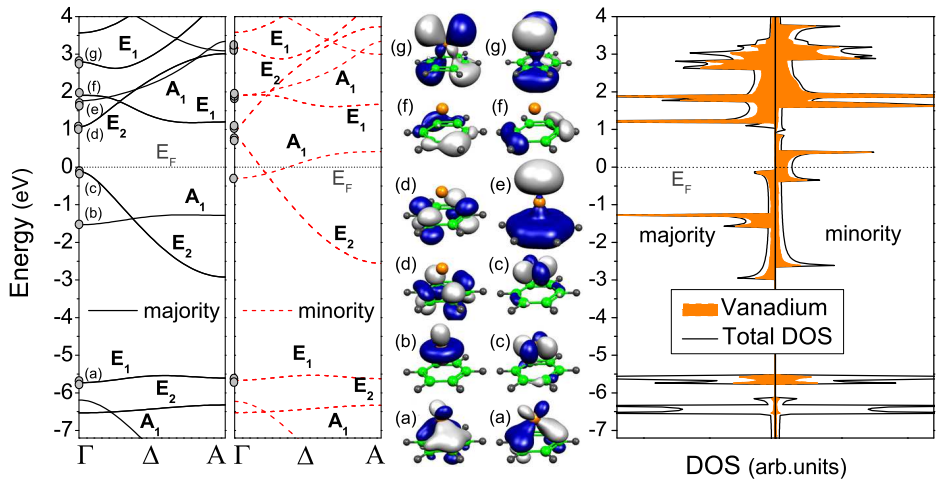}
   \end{minipage}
   \begin{minipage}[c]{0.21\textwidth}
\caption{\label{BAND} (Color online) The spin-resolved band
structure (left plot) and the DOS (right plot) of the
V(C$_6$H$_6$) wire in the ferromagnetic phase. The labels nearby
the circles at the band structure plot indicate crystalline
orbitals of the wire calculated for the $\Gamma$ point. These
functions are shown in the center. The corresponding orbitals of
the minority and majority spin bands have practically the same
shape. }
\end{minipage}
\end{figure*}

Our calculations are based on density functional theory and were
performed within the linear combination of atomic orbitals
formalism implemented in the highly accurate CRYSTAL code
\cite{9}. The Perdew-Wang \cite{10} parametrization for the
exchange and correlation energy was used. The description of the
electron subsystem was done with the full electron basis set
6-311G** \cite{11} for carbon and hydrogen atoms and with a frozen
core basis set SBKJC \cite{12} for vanadium. The convergent results
were achieved using the grid of 24 $k$-points in the 1D Brillouin zone.
The wire under consideration (Fig.1) has a
one-dimensional hexagonal lattice with $P6/mmm$ ($D_{6h}^1$) rod
symmetry. From calculations of the total energy of the system we
have found that the optimal structure corresponds to the following
interatomic distances: $d_{\mathrm{VBz}}$=1.69~\AA,
$r_{\mathrm{CC}}$=1.45~\AA~and $r_{\mathrm{CH}}$=1.09~\AA.
These data are in a good agreement with other
theoretical studies of V-Bz clusters \cite{16,17}. We have also
examined another, staggered-type, geometry of the wire when the
consecutive benzol rings were rotated by $30^{\circ}$ with respect
to each other. We have found that the staggered geometry is
energetically less favorable, with a total energy being 0.06~eV
higher than the lowest-energy of the wire with of $D_{6h}^1$
symmetry. Moreover, our calculations showed that the rotation of
the benzene ring does not change the electronic properties of the wire.

In Fig.2 we present the spin-polarized band structure of the V-Bz
wire (left plot) together with the corresponding density of states
(right plot). We accompanied these data by plotting the
crystalline orbitals evaluated at the $\Gamma$-point. The symmetry
of the bands corresponds to the $C_{6v}$ point group relevant for
the wire $k$-vectors. The notations of Ref.\cite{Group_Theory} are
used throughout the text. We see from Fig.2 that for the minority
spin electrons the $A_1$ band and the double degenerate $E_2$ band
cross the Fermi level, while for the majority spin electrons the
Fermi energy lies in the gap. The direct gap has a width of 
1.18~eV while the spin-flip gap is 0.12~eV at the $\Gamma$ point. Such
system would have ballistic conductance of $3\,e^2/h$ in one spin
channel and zero transmission for the other spin. To verify the
obtained result we have performed additional calculations within
the Vosko-Wilk-Nusair LDA-type \cite{13} and hybrid B3LYP
\cite{B3LYP} exchange-correlation potentials together with all
electron \cite{Ruiz} and frozen core LANL2DZ \cite{14} basis
sets. We have confirmed that the half-metallic behavior is robust
against details of the calculations. The exact value of the gap
width and relative positions of the bands are of course subject to
the chosen functional. The LDA predicts a bit smaller value for
the gap (around 1.0~eV) while the hybrid B3LYP functional
noticeably pulls the bands apart the Fermi level. We have also
examined the antiferromagnetic (AFM) state of the wire and found
it to be energetically less preferable as compared with the
ferromagnetic configuration.

We show in Fig.3 the total valence charge density
($\rho_{\uparrow}+\rho_{\downarrow}$) and spin density
($\rho_{\uparrow}-\rho_{\downarrow}$) contour plots for the
ferromagnetic V-Bz wire. The density plots are given for two
planes: along the wire and perpendicular to it. The electronic
charge density in the V-Bz bond critical point is 0.132
$e/{a_0^3}$. According to the Bader analysis \cite{18} of the
topology of the charge density, the covalent type of bonding is
predominant. The vanadium atom is charged negatively ($-0.22 e$)
due to a charge transfer to benzene that agrees with the findings
of Refs.\cite{6,19}. Thus, the ionic type of bonding is also
present. The magnetic density map (Fig.3b) indicates a quite
localized positive magnetic moment at the V atom (+1.28$\mu_B$)
and a small negative magnetic moment ($-0.28\mu_B$) redistributed
over six carbon atoms. Our value of the magnetic moment at the V
atom agrees well with the theoretical data of Ref.\cite{6}
obtained for finite V$_n$Bz$_{n+1}$ complexes where these values
varied from 1.15$\mu_B$ up to 1.36$\mu_B$ depending on the complex
size and the vanadium position in the complex. The total magnetic
moment of the wire unit cell is equal to 1.0~$\mu_B$. The same
integer value of the total magnetic moment was obtained in EPR
measurements \cite{20,21,22} as well as in the calculations
performed for the single V-Bz and multidecker V$_n$Bz$_{n+1}$
clusters with $n\leq6$ \cite{6,Muhida}.

\begin{figure*}
   \subfigure[\mbox{}]{
   \begin{minipage}[b]{0.49\textwidth}
      \centering \includegraphics[height=2.0in]{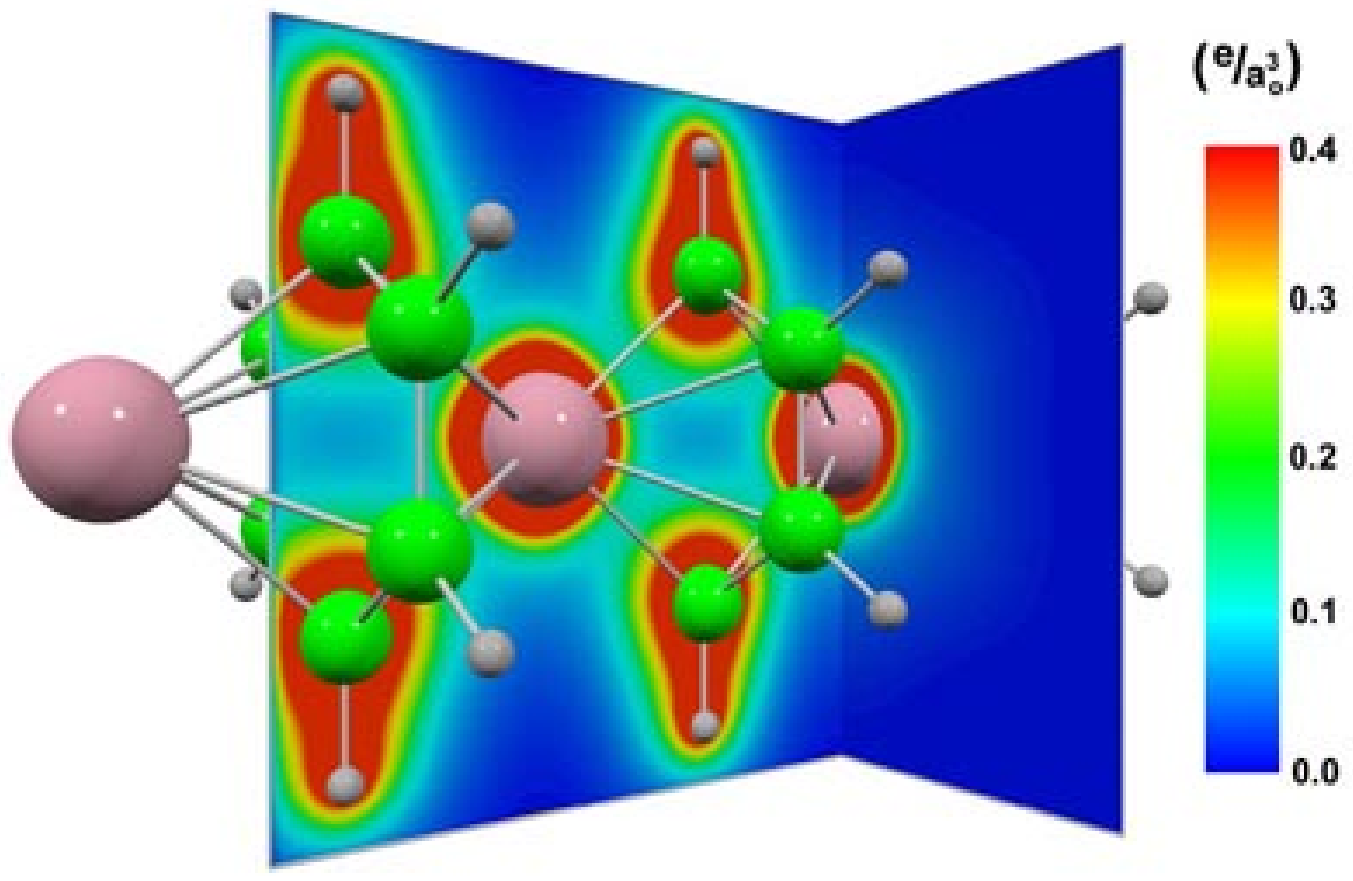}
      \label{DensTot}
   \end{minipage}}
   \subfigure[\mbox{}]{
   \begin{minipage}[b]{0.49\textwidth}
      \centering \includegraphics[height=2.0in]{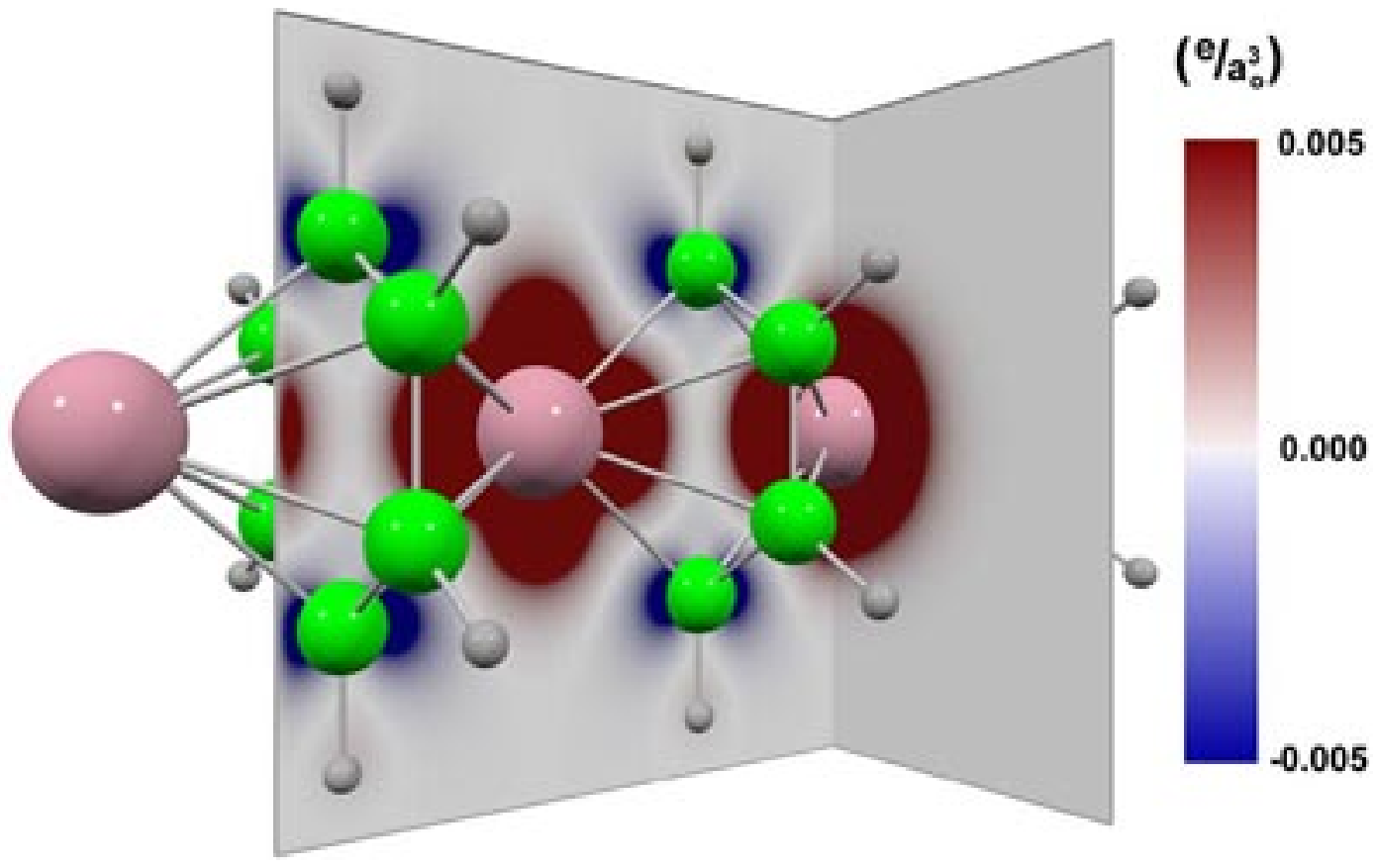}
      \label{DensMag}
   \end{minipage}}
\caption{\label{Dens} (Color) Total valence charge density (a) and spin
 density (b) for V(C$_6$H$_6$) wire.}
\end{figure*}

The analysis of the bands shown in Fig.2 together with data
presented in Fig.3 leads to the following picture of the formation
of the ground state. The fully occupied bands below $-6$~eV with
respect to $E_F$ arise from the to pure benzene orbitals which are
not spin-splitted and thus play no role in the formation of the
magnetic state. For energies above $-6$~eV the wire bands are
formed from the hybridized states of the vanadium and the benzene,
i.e.\ from molecular orbitals of the V-Bz cluster. The strong
crystalline field splits the vanadium $3d$ states to the singlet
$A_1$ state ($d_{3z^2-r^2}$) and two doublets, of $E_1$ ($d_{xz}$,
$d_{yz}$) and $E_2$ ($d_{xy},d_{x^2-y^2}$) symmetry. The doubly
degenerate $E_1$ band just above $-6$~eV is formed mainly from the
HOMO and HOMO$-1$ $\pi$-type orbitals of benzene with an admixture
of $d_{xz}$, $d_{yz}$ vanadium states. Due to the strong
hybridization effect between the states of the same symmetry, the
vanadium $d_{xz}$,  $d_{yz}$ levels are pushed well above the
Fermi energy where they are coupled with antibonding benzene
states leading to the formation of the two different bands of
$E_1$ symmetry, marked by (f) and (g) in Fig.2.  The remaining
$d_{3z^2-r^2}$ and $d_{xy}, d_{x^2-y^2}$ vanadium states form two
bands nearby the Fermi level, of $A_1$ and $E_2$ symmetry,
respectively [they are marked by (b) and (c)]. These bands are
spin-split. The vanadium atom itself has three $3d$ electrons. It
turns out that in the wire the vanadium $s$ states are shifted
above $E_F$ and are responsible for the formation of the $A_1$
band labelled as (e) in Fig.2. Thus, two $s$ electrons of vanadium
move to the $3d$ shell and five electrons in total wish to occupy
three levels ($A_1$ and $E_2$) which are available per spin.
Finally, majority spin electrons fill completely two bands of
$A_1$ and $E_2$ symmetry which therefore are placed below~$E_F$.
The remaining two electrons of each unit cell are redistributed
among minority spin $A_1$ and $E_2$ bands both of which are
crossing the Fermi level. To conclude, the magnetic moment per
unit cell is forced to be 1.0$\mu_B$. The semiconductor gap in the
majority spin channel is formed between two different $E_2$ bands
one of which originates mainly from $d_{xy}, d_{x^2-y^2}$ vanadium
states while the upper one comes from LUMO and LUMO$-1$
antibonding $\pi^{*}$-states of benzene.

Let us move again to Fig.3 and consider the spin density plot. We
see from Fig.2 that the $E_2$ minority spin states nearby $E_F$
are almost filled, therefore the magnetic moment at the vanadium
atom arises mainly from the $A_1$ states [labelled as (b)]. As it
is seen from Fig.2 the corresponding wave function (b) is nearly a
pure $d_{3z^2 - r^2}$ vanadium orbital so that its symmetry is
reflected in the angular distribution of the spin density around
the V atom. We note further more, that the small negative magnetic
moment on the C atoms results from the uncompensated $E_2$ band
(c) whose molecular orbital contains an admixture of the carbon
$p$ states (Fig.2).

\begin{figure}[b]
  \centering \includegraphics[bb=21 15 310 210, height=2.2in]{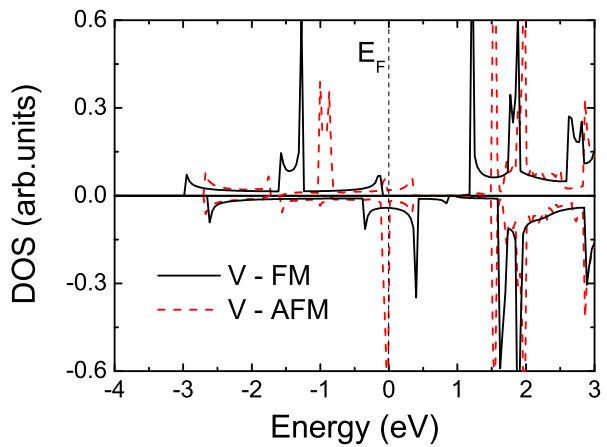}
  \caption{\label{DOS_F_AF}(Color online) Local density of states at the V atom for FM and AFM
   configurations of the wire (in the AFM state the second V atom has inverted
   LDOS concerning spin direction compared to the first one). }
\end{figure}

We can ask further the following question: what is the origin of
the exchange interaction between quite localized magnetic moments
of V atoms? According to the structure of the wire, we can assume
that different mechanisms of the exchange are possible: direct
exchange as well as the indirect, superexchange interaction which
in the present case could be established through the $\pi$-type
orbitals of benzene. To demonstrate that direct exchange between
vanadium moments indeed takes place we compare the spin-projected
local density of states (LDOS) at the V atom for the ferromagnetic
(FM) and the antiferromagnetic (AFM) state (see
Fig.\ref{DOS_F_AF}). If there would be no mutual influence of
localized moments, one could expect the LDOS for both phases to be
nearly indentical. As it is clearly seen from Fig.4, this is not
the case for the V-Bz wire. The structure of the LDOS changes for
the AFM phase. The AFM phase is not any more the half-metal and is
characterized by the reduced magnetic moment at the V atom
(0.62$\mu_B$). The direct exchange interaction between V moments
favors the FM phase to be energetically preferable. However, we
can not exclude another, indirect type of spin coupling, which is
governed by the Kramers-Anderson mechanism
\cite{Kramers_Anderson,Goodenough_Kanamori} of the superexchange
interaction. For the considered system, because of the overlap of
the HOMO (HOMO$-1$) $\pi$-type orbital of benzene with the
$d_{xz}$, $d_{yz}$ orbitals of vanadium, one of the $p$ electrons
from benzene can hop to one of the V atoms and the remaining
unpaired $p$ electron on Bz can enter into a direct exchange with
the other V atom. Applying the semi-empirical Goodenough-Kanamori
rules \cite{Goodenough_Kanamori} regarding to the sign of the
effective exchange integral, we can assume that the indirect
exchange interaction tends to aline vanadium moments antiparallel
and thus it competes with the direct coupling.

\begin{figure}[t]
  \centering
  \includegraphics[bb=21 15 310 210, height=2.0in]{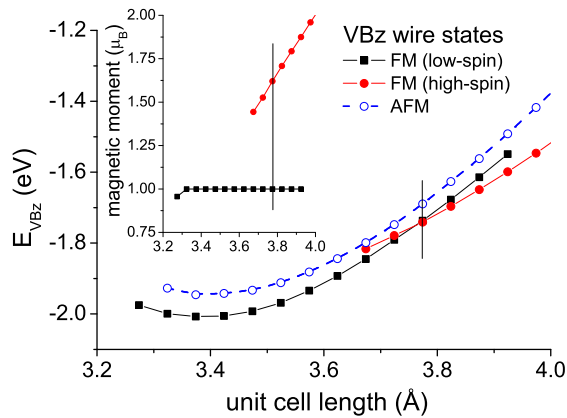}
  \caption{(Color online) Outer plot: Binding energy $E_{\mathrm{VBz}}$
   between V and Bz as a function of the unit cell length.
   Black and red lines correspond to the low- and high-spin FM states, respectively.
   Blue line with open circles refers to the AFM state which for any
lattice constants is less preferable as compared with FM state.
Innner plot: Behavior of the magnetic moment per unit cell under
elongation of the wire. }
\end{figure}

We have also studied the effect of the longitudinal stretching on
the properties of the VBz wire. Our results are summarized in
Fig.5 where we show binding energy between V and Bz as a function
of the unit cell length (outer plot) together with the behavior of
magnetic moment (innner plot).  We have found that a half-metallic
state characterized by the integer magnetic moment is conserved up
to $\sim 12\%$, elongation of the wire. At the critical unit cell
length of about 3.75~\AA, the system exhibits transition to the
high-spin state. Just before the critical point the partially
occupied minority spin $A_1$ band (b) nearby $E_F$ is transformed
to the very narrow peak pinned at the Fermi level. Under further
stretching this band is pushed above $E_F$ and a fractional charge
of about 0.3$e$ per unit cell from the minority spin band flows to
the upper lying majority spin band of $E_1$ symmetry (f) which now
intersects the Fermi level close to the A point of the Brillouin
zone. Thus, the transition is accompanied by an abrupt jump in the
magnetic moment (inserted in Fig.5), from 1$\mu_B$ up to $\sim
1.6\mu_B$ per unit cell, and high spin state becomes energetically
favorable because of the gain in the exchange~energy.

To conclude, using the {\it ab inito} density functional theory
based calculations we predict, at least theoretically, a new class
of half-metal ferromagnets: infinite one-dimensional
organometallic vanadium-benzene [V$_n$(C$_6$H$_6$)$_{n+1}$ ($n \to
\infty$)] wires. We mention, that according to our calculations,
the half-metallic properties have been also found for similar
systems, where V was substituted by Mn or Co as well as for hybrid
systems (M$_1$-Bz-M$_2$-Bz)$_n$ ($n \to \infty$), where M$_1$ and
M$_2$ stand for different (V, Mn or Co) atoms.

We would like to thank B.~Yavorsky, V.~Stepanyuk and F.~Evers for
fruitful discussions. This work was supported by DFG, SPP 1165
"Nanodr{\"a}hte und Nanor{\"o}hren".

\end{document}